\documentstyle[aps,preprint]{revtex}
\begin{document}
\setcounter{page}{0}
\def\footnoterule{\kern-3pt \hrule width\hsize \kern3pt}
\tighten
\title{Higher Derivative CP($N$) Model and Quantization of the
Induced Chern-Simons Term}

\author{ Taichi Itoh$^{1,2}$\footnote{Email address: taichi@knu.ac.kr}
and Phillial Oh$^1$\footnote{Email address: ploh@dirac.skku.ac.kr}
}
\address{
$^1$Department of Physics and Institute of Basic Science, 
Sungkyunkwan University\\Suwon 440-746, Korea\\
$^2$Department of Physics, Kyungpook National University, 
Taegu 702-701, Korea
}

\date{SKKUPT-02/2000, June 2000}

\maketitle
\draft

\begin{abstract}
We consider higher derivative CP($N$) model in $2+1$ dimensions with the
Wess-Zumino-Witten term and the topological current density squared term.
We quantize the theory by using the auxiliary gauge field formulation
in the path integral method and prove that the extended model remains
renormalizable in the large $N$ limit. We find that the Maxwell-Chern-Simons 
theory is dynamically induced in the large $N$ effective action at a 
nontrivial UV fixed point.  The quantization of the  Chern-Simons term is 
also discussed.
\end{abstract}



\newpage


The CP($N$) model in $2+1$ dimensions has many remarkable
properties. In contrast to the perturbative nonrenormalizability,
the theory is renormalizable in the large $N$ limit, in spite of the
appearance of linear divergence \cite{aref}.
It also exhibits a nontrivial fixed point structure which divides the
symmetric and broken phases with a second order phase transition,
and dynamical generation of a gauge boson \cite{coleman}.

The purpose of this article is to extend the CP($N$) model with higher
order derivative terms, and discuss its possible consequence in the
large $N$ limit. In general, adding higher derivative terms in the
nonlinear sigma models would make the theory less renormalizable and
the path integral more involved. However, with a specific
choice of higher derivative terms to be described below,
the theory admits the auxiliary gauge field formulation and an exact 
evaluation of the path integral can be performed in the large $N$ limit.
We find that (I) the extended theory remains renormalizable and (II)
the auxiliary gauge field becomes dynamical with induced
Maxwell-Chern-Simons terms.
In order to achieve these results, two types of higher derivative terms
are required. One is the third order derivative Wess-Zumino-Witten (WZW)
term which lives in $M_4$ whose boundary is our $2+1$
dimensional space-time.  It is well known that the coefficient of this term 
must be quantized in the case of Grassmann coset space ${\rm Gr}(N,n)\equiv
{\rm SU}(N)/{\rm SU}(N-n)\times{\rm U}(n)$ with $n> 1$ \cite{raje}. 
The path integral evaluation of this Grassmann model is very involved
even in the large $N$ limit due to the non-Abelian ${\rm U}(n)$
structure, but one can truncate the theory to the Abelian ${\rm U}(1)$
sector which yields the CP($N$) model. In the following, we consider
the CP($N$) model as  an Abelian truncated version of the Grassmann model, 
and a priori quantization of WZW coefficient survives.
This leads to the interesting consequence that the coefficient of the 
induced Chern-Simons term \cite{jack,jack1} is also quantized. 
The other is the topological current density squared term which was 
considered recently in the higher derivative extension of CP($N$) model 
in $1+1$ dimensions \cite{itoh}. 


Let us start directly from the auxiliary gauge field formulation of the
extended CP($N$) model for economic presentation.
We will shortly show that this theory is equivalent to the aforementioned
higher derivative CP($N$) model. The Lagrangian is given by
\begin{equation}
{\cal L}_A =\frac{N}{G}\left[(D_\mu z)^\dagger (D^\mu z)
-i\theta G A_\mu \,\epsilon^{\mu\nu\rho}(\partial_\nu z)^\dagger
(\partial_\rho z)-\lambda\,(z^\dagger z -1) \right],
\label{ac}
\end{equation}
where $D_\mu \equiv \partial_\mu -iA_\mu$ and $z$ is an $N$ components
complex scalar field which obeys a constraint $z^\dagger z=1$.
The first term is the usual 
${\rm CP}(N)\equiv {\rm SU}(N)/{\rm SU}(N-1)\times{\rm U}(1)$ model
in the auxiliary gauge field formulation. The second term will be
responsible for higher derivative terms when the auxiliary gauge field
$A_\mu$ are eliminated through the equations of motion.
Note that $z$ with the constraint $z^\dagger z=1$ contains $2N-1$ real
scalars, whereas the coset space is a ($2N-2$)-dimensional manifold.
This mismatch is due to the local U(1) symmetry of the model.
More specifically, the U(1) gauge transformation is given by
\begin{equation}
z(x)\to e^{i\alpha(x)}z(x),\quad
A_\mu (x)\to A_\mu (x) +\partial_\mu \alpha(x).
\end{equation}
Note that the first term in the Lagrangian (\ref{ac}) is manifestly
gauge invariant, whereas the second term changes by a total derivative,
hence the action is gauge invariant.

The field $z\equiv(z_1,\dots,z_N)^T$ is separated into $2N-2$
Nambu-Goldstone bosons $\psi\equiv(z_1,\dots,z_{N-1})^T$ associated with
the spontaneously broken SU($N$) symmetry and Higgs bosons
$z_N \equiv (\sigma+i\chi)/\sqrt{2}$.
In general there are two possible phases \cite{coleman}:
(I) $\langle\sigma\rangle\neq 0$, $\langle\lambda\rangle=0$ and
(II) $\langle\sigma\rangle=0$, $\langle\lambda\rangle\neq 0$.
In phase (I) both global SU($N$) and local U(1) symmetries are broken
simultaneously and $\psi$ arise as massless Goldstone bosons.
Through the Higgs mechanism $\chi$ turns to a longitudinal mode of
massive gauge boson $A_\mu$. On the other hand in phase (II) both
global SU($N$) and local U(1) symmetries are not spontaneously broken.
Instead $\psi$ and $z_N$ are combined into $z$ with a universal mass
$\langle\lambda\rangle^{1/2}$.
We will see later that the dimensionless coupling $u\equiv G\Lambda$
shows a nontrivial ultraviolet (UV) fixed point $u^*$ which arises
as a zero of the Callan-Symanzik $\beta$-function and separates the weak
coupling broken phase (I) from the strong coupling symmetric phase (II).
Since we are interested in dynamical generation of gauge bosons, we
confine our computation to the symmetric phase alone.

Solving the equations of motion and eliminating the $A_\mu$ fields,
we see that the Lagrangian becomes
\begin{equation}
{\cal L}_J =\frac{N}{G}\Biggl[(\partial_\mu z)^\dagger (\partial^\mu z)
-J_\mu J^\mu +\theta G\,\epsilon^{\mu\nu\rho} J_\mu \partial_\nu J_\rho
+\frac{1}{4}\theta^2 G^2 J_\mu (\partial^2 g^{\mu\nu} -\partial^\mu
\partial^\nu) J_\nu \Biggr],
\label{lagj}
\end{equation}
where $J^\mu \equiv (1/2i)[z^\dagger \partial^\mu z
-(\partial^\mu z^\dagger)z]$ with the constraint $z^\dagger z=1$.
We see that the extra third and fourth terms as well as the original
second term are not renormalizable in perturbative expansion.
The geometrical implication of the above Lagrangian can be seen in the
coadjoint orbit approach for the nonlinear sigma model \cite{oh,bal}.
In terms of the coadjoint orbit variable
\begin{equation}
Q=-iz z^\dagger+i\frac{I}{N},~~ z^\dagger z=1,
\label{cpn}
\end{equation}
and the topological current density $t^\mu=\epsilon^{\mu\nu\rho}
\!<\!Q\,\partial_\nu Q\,\partial_\rho Q\!>\,
=-\epsilon^{\mu\nu\rho}\,\partial_\nu J_\rho$,
we find that the Lagrangian (\ref{lagj}) is equivalent to
\begin{eqnarray}
S_Q=-\frac{N}{G}\int d^3x\, \Biggl[\frac{1}{2}\!<\!\partial_\mu
Q\,\partial^\mu Q\!> +\frac{1}{4}\,\theta^2 G^2\,t^\mu t_\mu
\Biggr]+S_{WZW}.
\label{qac}
\end{eqnarray}
Here the symbol $< \cdots >$ stands for a  trace over an $N\times N$
matrix. And $S_{WZW}$ descends from $M_4$ whose boundary is 
our $2+1$ dimensional space-time, and is given by
\begin{equation}
S_{WZW}= -iN\theta \int _{M_4}\!\!d^4 x\,\epsilon^{\mu\nu\rho\sigma}
\!<\!Q\, \partial_\mu Q\, \partial_\nu Q\, \partial_\rho
Q\, \partial_\sigma Q\!>,
\label{wzw}
\end{equation}
where $N\theta$ has to be $k/4\pi$ for integer $k$ 
from the aforementioned quantization condition \cite{raje}.
Note that the gauge formulation requires the coefficients of the
topological current squared term to be fixed in terms of $\theta$.

The equivalence between (\ref{lagj}) and (\ref{qac}) can be conveniently 
checked in terms of differential forms \cite{eguchi}.
Let us introduce 1-forms $dQ = (\partial_\mu Q)dx^\mu$ and
$J\equiv J_\mu dx^\mu=-iz^\dagger dz=idz^\dagger z$ with  
$z^\dagger z=1$. Then, by the Hodge dual $*$, we have the 1-form of 
topological current density given by $t\equiv t_\mu dx^\mu =-*dJ$.
Now, using $i dQ = dz z^\dagger +z\, dz^\dagger$, the first term in
(\ref{qac}) can be rewritten as 
\begin{eqnarray}
-\frac{1}{2}\int d^3 x <\!\partial_\mu Q\,\partial^\mu Q\!>\,\,=
\frac{1}{2}\int\! <\!*dQ\wedge dQ \!>\,\,=
-\int \left( *dz^\dagger \wedge dz -*J\wedge J\right),
\end{eqnarray}
which reproduces the first two terms in (\ref{lagj}). 
Similarly the second term in (\ref{qac}) can be shown to be equal to
\begin{eqnarray}
-\int d^3 x~t_\mu t^\mu=
\int (*t \wedge t) =
\int (dJ\wedge *dJ)=\int \!d^3 x\, \epsilon_{\alpha\beta\gamma}
\epsilon^{\mu\nu\gamma}(\partial^\alpha J^\beta)(\partial_\mu J_\nu),
\end{eqnarray}
which corresponds to the fourth term in (\ref{lagj}) after a partial 
integration. 
Finally, WZW action (6), which is an integration over the $M_4$ of 
the 4-form $<\!Q(dQ)^4 \!>$, yields after a straightforward computation 
\begin{eqnarray}
\int_{M_4}\! <\!Q(dQ)^4\!>\,\,= -i \int_{M_4}
\left(dz^\dagger \wedge dz \wedge dz^\dagger \wedge dz\right) 
=i \int_{M_4}\! d( J\wedge dJ),
\end{eqnarray}
where we have used $\int (z^\dagger dz)^4= 0 =
\int (dz^\dagger \wedge dz)\wedge (z^\dagger dz)^2$ repeatedly.
It yields the third term in (\ref{lagj}) using the Stokes' theorem.


The original CP($N$) model is not renormalizable in larger than two
dimensions. However the theory may become renormalizable
through a resummation of Feynman diagrams in a different way from the
coupling perturbation expansion. In fact the $1/N$ expansion provides
such a resummation technique and it makes the CP($N$) model in less than
four dimensions renormalizable\cite{aref,rose}.

We can rewrite the Lagrangian (\ref{ac}) up to total derivative terms as
\begin{equation}
{\cal L}_{A}^{\prime}=\frac{N}{G}\,z^\dagger\left[-\,\partial^2 -m^2
-\Gamma\right]z+\frac{N\lambda}{G},
\end{equation}
where we separate the Goldstone boson mass $m^2$ from
$\lambda\equiv m^2 +\tilde{\lambda}$ and $\Gamma$ stands for the
interaction vertices:
\begin{equation}
\Gamma\equiv\tilde{\lambda}
-iA_\mu (\partial^\mu -\overleftarrow{\partial^\mu}
-\theta G\,\epsilon^{\mu\nu\rho}\overleftarrow{\partial_\nu}\partial_\rho)
-A_\mu A^\mu,
\end{equation}
where $\overleftarrow{\partial^\mu}$ and $\overleftarrow{\partial_\nu}$
do not operate on $A_\mu$. Path integrating $z$ and $z^\dagger$ provides
the large $N$ effective action:
\begin{equation}
S_{\rm eff}=\int\!d^3 x\,{\cal L}_A^{\prime} +iN\,{\rm Tr}\,{\rm Ln}\!
\left[-\,\partial^2 -m^2\right]-iN\sum_{n=1}^{\infty}
\frac{1}{n}\,{\rm Tr}\!\left[\frac{1}{-\,\partial^2 -m^2}\,
\Gamma\right]^n.
\end{equation}
We divide the effective action up to quadratic terms ($n=1,2$) into two 
parts, $S_{\rm eff}=S^I +S^{II}$
where $S^I$ denotes the large $N$ effective action in the original
CP($N$) model and $S^{II}$ stands for the extra induced terms.
After some straightforward calculations, we obtain
\begin{eqnarray}
S^I&=&N\int\!d^3 x\,\Biggl[\frac{1}{G}\,
z^\dagger\left[-\,\partial^2 -m^2 -\Gamma^I\right]z
+\left(\frac{1}{u}-\frac{1}{u^*}\right)\!\Lambda(m^2 +\tilde{\lambda})
\nonumber\\ &&
+\frac{1}{4\pi}m\tilde{\lambda}+\frac{1}{6\pi}m^3
+\frac{1}{2}\,\tilde{\lambda}\,\Pi_{\lambda}(i\partial)\,\tilde{\lambda}
-\frac{1}{4}\,F_{\mu\nu}\,\Pi_1 (i\partial)\,F^{\mu\nu} \Biggr],
\label{Seff2}\\
S^{II}&=&N\int\!d^3 x\,\Biggl[iA_\mu \,\epsilon^{\mu\nu\rho}
(\partial_\nu z)^\dagger (\partial_\rho z)
-\frac{1}{12}\,F_{\mu\nu}\,\theta^2 G^2 \,\Pi_2(i\partial)\,F^{\mu\nu}
\nonumber\\ &&
-\frac{2}{3}\,A_\mu \,\theta G\,\Pi_2 (i\partial)\,
\epsilon^{\mu\nu\rho}\,\partial_\nu A_\rho \Biggr],\label{Seff3}
\end{eqnarray}
where we have introduced the dimensionless coupling $u\equiv G\Lambda$
and $u^* \equiv 2\pi^2$. $\Gamma^I$ is the interaction vertices in
the original CP($N$) model without higher derivative terms,
and the vacuum polarization functions are given by
\begin{eqnarray}
\Pi_{\lambda}(p)&=&
\frac{1}{4\pi\sqrt{-p^2}}\arctan\frac{\sqrt{-p^2}}{2m},\\
\Pi_1 (p)&=&
\frac{1}{4\pi p^2}\left[m-\frac{4m^2 -p^2}{2\sqrt{-p^2}}
\arctan\frac{\sqrt{-p^2}}{2m}\,\right],\\
\Pi_2 (p)&=&\frac{1}{2\pi^2}\,\Lambda-\frac{3}{8\pi}\,m
+\frac{3}{4}\,p^2 \Pi_1 (p).\label{exvac}
\end{eqnarray}
We realize that there arise linear divergences in induced Chern-Simons and
Maxwell terms which have no counter terms in the original Lagrangian.


Renormalization of the coupling $G$ can be worked out in the same manner
as in the original CP($N$) model. The large $N$ effective potential
is defined as the effective action divided by $\Omega\equiv\int d^3 x$
with $\tilde{\lambda}\equiv0$, $A_\mu \equiv0$ and $z(z^\dagger)\equiv 0$.
It is given by
\begin{equation}
\frac{1}{N}V_{\rm eff}=-\left(\frac{1}{u}-\frac{1}{u^*}\right)
\!\Lambda m^2 -\frac{m^3}{6\pi}.
\end{equation}
The Goldstone boson mass $m$ is determined as
a nontrivial solution to the gap equation $dV_{\rm eff}/dm^2=0$
and reads
\begin{equation}
m =4\pi\Lambda\!\left(\frac{1}{u^*}-\frac{1}{u}\right).
\label{gapsol}
\end{equation}
We notice that $m$ can be independent of the ultraviolet cutoff $\Lambda$
by imposing $\Lambda$ dependence on the coupling $u$.
In fact the scale invariance condition $\Lambda dm/d\Lambda=0$ leads us
to the Callan-Symanzik $\beta$-function
\begin{equation}
\beta(u)\equiv\Lambda\frac{du}{d\Lambda}=u\!\left(1-\frac{u}{u^*}\right),
\label{beta}
\end{equation}
which shows a nontrivial UV fixed point at $u=u^*$.
In the original CP($N$) model the only divergence is the one which arises
in the large $N$ effective action through a tadpole diagram coupled with
$\tilde{\lambda}$ so that the scale invariance condition
$\Lambda dm/d\Lambda=0$
is enough to achieve the cutoff independent theory.
Since $m$ is scale invariant, the solution to the gap equation (\ref{gapsol})
suggests that the renormalization of coupling is given by
\begin{equation}
\Biggl(\frac{1}{u}-\frac{1}{u^*}\Biggr)\Lambda=
\Biggl(\frac{1}{u_R}-\frac{1}{u_R^*}\Biggr)\mu,
\label{gr}
\end{equation}
where $u_R$ is the renormalized coupling at a reference energy scale $\mu$.

In the extended model, however, linear divergences arise in the
induced Chern-Simons and Maxwell terms which do not have their
counter terms in the classical action. Therefore the higher derivative 
theory seems to be nonrenormalizable, although the coupling $u$ can be 
renormalized in the same way as in the original CP($N$) model.
However, since the extra linear divergences are always accompanied by
the coupling $G\equiv u/\Lambda$ which cancels the linear divergences,
we expect the large $N$ effective action (\ref{Seff3}) to be scale
invariant in the continuum limit $\Lambda\to\infty$.

To study this point in more detail, we first look at how $S^I$ can be
scale invariant through the renormalization procedure.
The induced kinetic terms of $A_\mu$ and $\tilde{\lambda}$ are UV finite
in themselves so that we do not need wave function renormalization for
them. Then the second term in the right hand side of Eq.\ (\ref{Seff2})
becomes UV finite through Eq.\ (\ref{gr}) from which the Z factor for
the coupling $G$ can be read
\begin{equation}
Z^{-1}\equiv\frac{G_R}{G}=1-\frac{u_R}{u_R^*}+\frac{u_R}{u^*}
\left(\frac{\Lambda}{\mu}\right).
\end{equation}
Here $G_R$ is connected to the dimensionless coupling $u_R$ through
$u_R \equiv G_R \mu$.
The kinetic term of $z$ has to be UV finite in itself so that we see
\begin{equation}
\frac{1}{G}\,z^\dagger\left[-\,\partial^2 -m^2 \right]z=
\frac{1}{G_R}\,z_R^\dagger\left[-\,\partial^2 -m^2 \right]z_R,
\label{zwave}
\end{equation}
where $z_R$ has been introduced through $z=Z_z^{1/2}z_R$ and $Z_z$ is
thereby determined as $Z_z \equiv Z$ in order to cancel the Z factor
from the coupling renormalization. Thus we realize that $\Gamma^I$
in $S^I$ has to remain invariant through the renormalization procedure.
This forces both of $A_\mu$, $\tilde{\lambda}$ to be unchanged through
renormalization. This is consistent with the UV finiteness of kinetic
terms for $A_\mu$ and $\tilde{\lambda}$ in $S^I$.

Now that we have explicitly shown that the original CP($N$) model
can be renormalized through the $1/N$ resummation, let us look at what
happens in $S^{II}$ if we take the continuum limit $\Lambda\to\infty$.
The Callan-Symanzik $\beta$-function (\ref{beta}) tells us that the
dimensionless coupling $u\equiv G\Lambda$ goes to the UV fixed point
$u=u^*$ as the cutoff $\Lambda$ goes to infinity, whereas the bare
coupling $G\equiv u/\Lambda$ reduces to zero. On the other hand,
the extra vacuum polarization effect $\Pi_2$ in Eq.\ (\ref{exvac})
can be rewritten as
\begin{equation}
G\,\Pi_2 (p) = \frac{u}{u^*}-G\left[\frac{3}{8\pi}\,m
-\frac{3}{4}\,p^2 \Pi_1 (p)\right].
\end{equation}
Therefore, we can obtain the UV finite result $G\,\Pi_2 (p)\to 1$
as $\Lambda\to\infty$. In the same reasoning the extra contribution
to the Maxwell term which contains $G^2\,\Pi_2 (p)$ vanishes in the
continuum limit. Moreover, the first term in $S^{II}$ has an extra $G$
after the renormalization (\ref{zwave}) and is thereby suppressed by
a factor $1/\Lambda$ as $\Lambda\to\infty$. Thus in the continuum limit
$S^{II}$ becomes a UV finite Chern-Simons action
$-(2N\theta/3)\int\!d^3 x\,\epsilon^{\mu\nu\rho}A_\mu \partial_\nu A_\rho$
without any ambiguity.

According to power counting of superficial degrees of UV divergences,
the three-points function of $A_\mu$ shows a linear divergence.
However the extra gauge coupling in $\Gamma$ is invariant under the
charge conjugation so that the three points function and its charge
conjugation cancel each other in the same way as in the original
CP($N$) model \cite{aref}. 
Moreover all $n$-points functions with $n\ge4$ are
UV finite and the contributions from the extra gauge interaction are
accompanied by $G$. Therefore in the continuum limit such extra effects
are suppressed by a factor $1/\Lambda$ and become irrelevant.
Hence we can conclude that the large $N$ effective action in our
extended model is renormalizable and the gauge sector is equivalent to
the Maxwell-Chern-Simons theory which couples minimally to $z$ field.
The effective Lagrangian at the lowest order in the derivative expansion
can be rewritten as
\begin{equation}
{\cal L}_{\rm eff}=-\frac{1}{4}\,{\cal F}_{\mu\nu} {\cal F}^{\mu\nu}
-\frac{\kappa}{2}\,\epsilon^{\mu\nu\rho}{\cal A}_\mu \partial_\nu
{\cal A}_\rho,
\end{equation}
where we introduced $e^2\equiv 24\pi m/N$ and redefined the gauge
field by $A_\mu =e{\cal A}_\mu$ so that the covariant derivative is
$D_\mu \equiv \partial_\mu -ie{\cal A}_\mu$.
In this Lagrangian, the Chern-Simons coefficient becomes a dimensionful 
parameter $\kappa\equiv32\pi m\theta$, and we supposed that
the Goldstone bosons became very massive and decoupled after the symmetry
restoration of both global ${\rm SU}(N)$ and local ${\rm U}(1)$. 

Note that the ratio of the topological gauge boson mass $\kappa$
\cite{jack1} to the effective gauge coupling square $e^2$ is proportional 
to $N\theta$ and is quantized such as $\kappa/e^2 =k/3\pi$.
Recently, some authors showed that in the Maxwell-Chern-Simons theory
coupled to fermion fields, the Lorentz symmetry is spontaneously broken
through a dynamically induced magnetic field when $\kappa/e^2$ is
quantized in a unit of $1/4\pi$ \cite{hoso,itoh2}.
Specifically, Ref.\ \cite{itoh2} introduced $N_f$-flavored four-components
fermion fields and showed that only the Lorentz symmetry is broken in
$\kappa=N_f e^2/2\pi$, whereas both flavor U$(2N_f)$ and Lorentz symmetries
are broken at the same time in $\kappa=N_f e^2/4\pi$ through a dynamically
generated fermion mass and a magnetic field. Our current results provides
a geometrical origin of quantization of the Chern-Simons coefficient
due to the WZW term. In fact the quantization condition $\kappa=N_f e^2/2\pi$
and  $N_f e^2/4\pi$ correspond to $2k=3N_f$ and  $4k=3N_f$, respectively.
We may also possibly prove that the induced Chern-Simons term
may not receive any higher order $1/N$ corrections \cite{seme}.


We have investigated the gauge formulation of higher derivative CP($N$) 
model in $2+1$ dimensions with WZW term and topological current density
squared, and have proved its renormalizability in the large $N$ limit.
We also have found that the Maxwell-Chern-Simons theory is dynamically
generated in the effective action and the coefficient of the induced
Chern-Simons term must be quantized which is a direct consequence of
the quantization of the WZW term.
If we couple the theory to fermions with U$(2N_f)$ flavors, the low energy
effective theory shows spontaneous break down of the Lorentz symmetry
associated with an induced magnetic field when $2k=3N_f$, $4k=3N_f$.
It would be interesting to check whether this has some physical
implications, especially in condensed matter phenomena
such as anyon physics and the fractional quantum Hall effect \cite{frad}.

There exist related subjects to be studied further.
One of them is to compute the current algebra associated with
(\ref{qac}) and (\ref{wzw}), and to check whether it has some special
properties at the fixed point $u=u^*$.
Another interesting problem is to consider possible extension to higher
dimensions. The theory in $2+1$ dimensions was special in the sense that
the topological current $t^\mu$ interacts with original gauge field of
CP($N$) model. In $D$ dimensional extension, however, one would need
extra $D-2$ rank antisymmetric tensor fields $B_{\mu_3\cdots\mu_D}$
with interaction
\begin{eqnarray}
i\epsilon^{\mu_1\mu_2\mu_3\cdots\mu_D}(\partial_{\mu_1} z)^\dagger
(\partial_{\mu_2} z) B_{\mu_3\cdots\mu_D}
+\frac{1}{2M^{D-4}}B_{\mu_3\cdots\mu_D}B^{\mu_3\cdots\mu_D}.
\end{eqnarray}
This theory which corresponds to the auxiliary field formulation of
the dual variable description of SU(2) Yang-Mills theory in the
infrared limit \cite{fadd} for CP(2) case, is not renormalizable
in $D\geq 4$ even in $1/N$ expansion.
So we have to consider it as an effective theory which
describes a massless gauge field interacting with massive $H=dB$ field
in the $B\,^*\!F$-type interaction. It remains to investigate whether 
this type of Maxwell-Kalb-Ramond theory \cite{kalb}
has some relevance with quark confinement in $3+1$ dimensions 
\cite{kond,poly}.
\\

T.I. was supported by KOSEF Postdoctoral Fellowship and Korea Research
Center for Theoretical Physics and Chemistry.
P.O. was supported by the Samsung Research Fund, Sungkyunkwan
University, 1999, and in part by the KOSEF through the project number
2000-1-11200-001-3. This work was also partially supported by BK21 Physics
Research Program.

\end{document}